\documentclass[runningheads]{llncs}
\usepackage[T1]{fontenc}
\usepackage{graphicx}
\begin{document}
\title{FSE Compensated Motion Correction for MRI Using Data Driven Methods}
%
%
\author{Brett Levac\thanks{Equal contribution}, Sidharth Kumar$^{\star}$, Sofia Kardonik, Jonathan I. Tamir}
\authorrunning{B. Levac et al.}
%

\institute{
 The University of Texas at Austin, Austin TX 78705, USA\\
 \email{\{blevac, sidharth.kumar, sofia.kardonik, jtamir\}@utexas.edu}}
\maketitle              
\begin{abstract}
Magnetic Resonance Imaging (MRI) is a widely used medical imaging modality boasting great soft tissue contrast without ionizing radiation, but unfortunately suffers from long acquisition times. Long scan times can lead to motion artifacts, for example due to bulk patient motion such as head movement and periodic motion produced by the heart or lungs. Motion artifacts can degrade image quality and in some cases render the scans nondiagnostic. 
To combat this problem, prospective and retrospective motion correction techniques have been introduced. More recently, data driven methods using deep neural networks have been proposed. 
As a large number of publicly available MRI datasets are based on Fast Spin Echo (FSE) sequences,  methods that use them for training should incorporate the correct FSE acquisition dynamics.
Unfortunately, when simulating training data, many approaches fail to generate accurate motion-corrupt images by neglecting the effects of the temporal ordering of the k-space lines as well as neglecting the signal decay throughout the FSE echo train.
In this work, we highlight this consequence and demonstrate a training method which correctly simulates the data acquisition process of FSE sequences with higher fidelity by including sample ordering and signal decay dynamics. 
Through numerical experiments, we show that accounting for the FSE acquisition leads to better motion correction performance during inference. 

\keywords{Motion Correction \and Deep Learning \and Magnetic Resonance Imaging \and Fast Spin Echo.}
\end{abstract}
\section{Introduction}
MRI is a powerful medical imaging modality due to its superb soft tissue contrast which can help in diagnosing various types of pathology.
However, unlike imaging methods that use ionizing radiation such as CT, MRI is predominantly slow, and each MRI scan can take several minutes to acquire. This is because the MRI acquisition involves repeated application of signal excitation through radio-frequency (RF) pulses and signal reception using spatially varying magnetic field gradients.  Due to physical limitations including signal decay, only a small number of measurements in k-space can be acquired in a single excitation, and a duration (called the repetition time, TR) must pass before the MRI signal resets to equilibrium or steady state. As a result, multiple excitations are needed to fully sample k-space. Patient motion during the scan therefore commonly occurs, and leads to degraded image quality \cite{slipsager2020}. If the degradation is not too severe the image might still be useful for clinical diagnosis. However, oftentimes, the motion is so severe that the image is nondiagnostic, and the scan needs to be repeated. It has been reported that motion-corrupted scans lead hospitals to incur nearly \$365K in additional costs each year \cite{slipsager2020}. 

Due to the high associated cost, many methods have been proposed to mitigate and correct for motion \cite{brain_motion_cor}. A common approach to reduce motion is to modify the acquisition and change the sampling pattern and ordering so that the motion artifacts are incoherent, as is the case for non-Cartesian radial sampling \cite{zaitsev2015motion}. Prospective motion correction during the scan is also possible using external sources such as respiratory bellows, nuclear magnetic resonance probes \cite{Pruessmann}, optical and electromagnetic tracking \cite{afacan2020retrospective}, and wireless RF tones \cite{PILOT}, or through self-navigators \cite{Butterfly}. However, prospective motion correction in some cases reduces scan efficiency, as motion-corrupt measurements must be re-acquired with additional RF excitations. Retrospective motion correction is also popular \cite{KURUGOL2017124,coll2021retrospective}, as it relies on computational processing during reconstruction in lieu of reducing scan efficiency.

Recently, data-driven approaches to retrospective motion correction have increased in popularity due to the rising success of deep learning and the availability of public MRI datasets with raw k-space \cite{zbontar2018fastmri}. These approaches are attractive as they do not require sequence modification, external hardware, or explicit knowledge of the motion information, and instead use a large corpus of training data to learn motion artifact removal. The motion removal can be posed as either supervised or self-supervised learning, in which a motion-corrupt image is ``denoised'' based on a motion-free reference. Many works that follow this paradigm simulate motion effects by corrupting k-space lines through linear phase augmentation and rotation \cite{johnson2019}.

When simulating motion artifacts for data driven motion correction, the ground truth k-space is divided into non-overlapping regions which are taken from different motion states of the underlying image. This structure of motion-corruption  reflects the fact that in FSE scans the time between phase encode lines in different repetition times (TR) (inter-TR time) is much longer than the time between echoes in a single TR (intra-TR time) and thus motion is roughly simulated as being inter-TR and not occurring between lines in the same echo train. Many of the publicly available datasets are acquired using fast spin-echo (FSE) imaging \cite{zbontar2018fastmri,BRATS,realnoiseMRI,mridata_org}, in which multiple k-space lines are rapidly acquired per RF excitation through the use of RF refocusing pulses. Therefore, the specific phase encode lines, their ordering, and signal decay modulation are all pre-determined by the acquisition parameters and the tissue relaxation parameters. In this setting it is often the case that when retrospectively applying motion to different lines of k-space the phase encode lines are grouped into non-overlapping subsets which are agnostic to the sampling pattern that was used to collect the data. This approach is fundamentally incorrect as it neglects the fact that signal decay is occurring within each echo train of the ground truth FSE image and thus training on such data will lead to inconsistent results when applying data driven methods to actual FSE data.

In this work, we propose a method for simulating synthetic motion artifacts that accounts for the echo ordering and signal decay present in FSE acquisitions, and we highlight the disparity in performance that  occurs when training on simulated data that is agnostic to the acquisition parameters.


\subsection{Related Works}
In the case of retrospective motion correction with unknown motion, both physics and learning based methods have been proposed. Authors in \cite{sense_encode} showed that by modeling the forward model of in-plane motion, an alternating minimization approach can be used to solve for both the parameterized motion model and the motion-free image. Similarly, authors in \cite{Tamer} proposed an algorithm to estimate the rotation and translation operator and used the estimated operator to remove the motion artifacts. Improving on that, the same authors used a convolutional network to first estimate the motion operator and then reconstruct the image with information and data consistency \cite{Namer}.   
The work in \cite{MC2-Net} used image alignment to remove the motion artifacts by inputting multiple images to the network among which one has motion artifacts. 

Authors in \cite{pawar2019suppressing} used an Inception-ResNet based encoder-decoder network to correct motion artifacts, the motion generation method only uses two randomly generated motion states.
Whereas \cite{singh2020joint} proposes a joint image and frequency space inter connected network to remove motion artifacts, the motion artifacts are simulated by drawing translation and rotation parameters from a uniform distribution. 
Authors in \cite{johnson2019} used a Conditional Generative Adversarial Network (GAN)  to correct motion-corrupt images which were generated using simulating translation and rotational motions. The considered motion model only used $2$ or $3$ motion events and the k-space data was acquired without considering the underlying scanning protocol.

\section{Methods}
\subsubsection{MR Forward Model with Rigid Body Motion:}
 Neglecting multiple coils and assuming Nyquist sampling, the forward model is as follows: 
\begin{equation}
\label{eqn:forward_model}
    y = Fx + w, \quad w\sim\mathcal{N}(0,\sigma^2),
\end{equation}
where $x$ is the image to be reconstructed, $F$ is the Fourier transform operator, $y$ represents the k-space measurements for a particular scan, and $w$ is additive white Gaussian noise. This model assumes that there is no motion during the scan. 
With the inclusion of motion, the forward model is as follows:
\begin{equation}
\label{eqn:motion_forward_model}
    y = PFM(x) + w, \quad w\sim\mathcal{N}(0,\sigma^2).
\end{equation}
Here, $P$ is a sampling operator which takes non-overlapping k-space samples from each TR, and $M(\cdot)$ contains the linear motion operation for each TR. If the inverse problem is solved assuming no motion ($M=I$), then the solution will be a motion-corrupt.

\subsubsection{MR Forward Model with Rigid Body Motion and Signal Decay:}
The forward model in (\ref{eqn:motion_forward_model}) assumes that k-space is acquired at the same echo time which is only true for steady-state and spin-echo sequences where a single k-space line is collected in each excitation. However, nearly all clinical MRI exams employ FSE sequences in which multiple echoes are acquired per excitation, and each echo has a decay factor depending on T2 relaxation and the time when the echo is acquired. This gives rise to the well-known T2-blurring seen in FSE imaging, and can be interpreted as k-space filtering leading to spatially varying convolutions in image-space \cite{T2_shuffling}. 
To account for this decay, the forward model is modified as follows:
\begin{equation}
\label{eqn:decay_motion_forward_model}
    y = PFM(h(TE, T2, PD )) + w, \quad w\sim\mathcal{N}(0,\sigma^2),
\end{equation}
where
\begin{equation}
\label{eqn:exp_model}
    h(TE, T2, PD ) = PD e^{\frac{-TE}{T2}}.
\end{equation}
Here $TE$ contains all the echo times at which the signal is being acquired, T2 is transverse relaxation time, and PD is the proton density. $h(\cdot)$ represents the exponential decay in the transverse magnetization as TE increases.
For simplicity we neglect the effects of imperfect RF refocusing pulses and we assume complete T1 recovery, i.e. long TR, though these  
 effects are easy to incorporate into the signal model \cite{T2_shuffling} due to the differentiability of the extended phase graph algorithm \cite{weigel_EPG}.


\subsection{Image Simulations}
\subsubsection{Ground Truth FSE Image:}
\begin{figure}
\includegraphics[width=\textwidth]{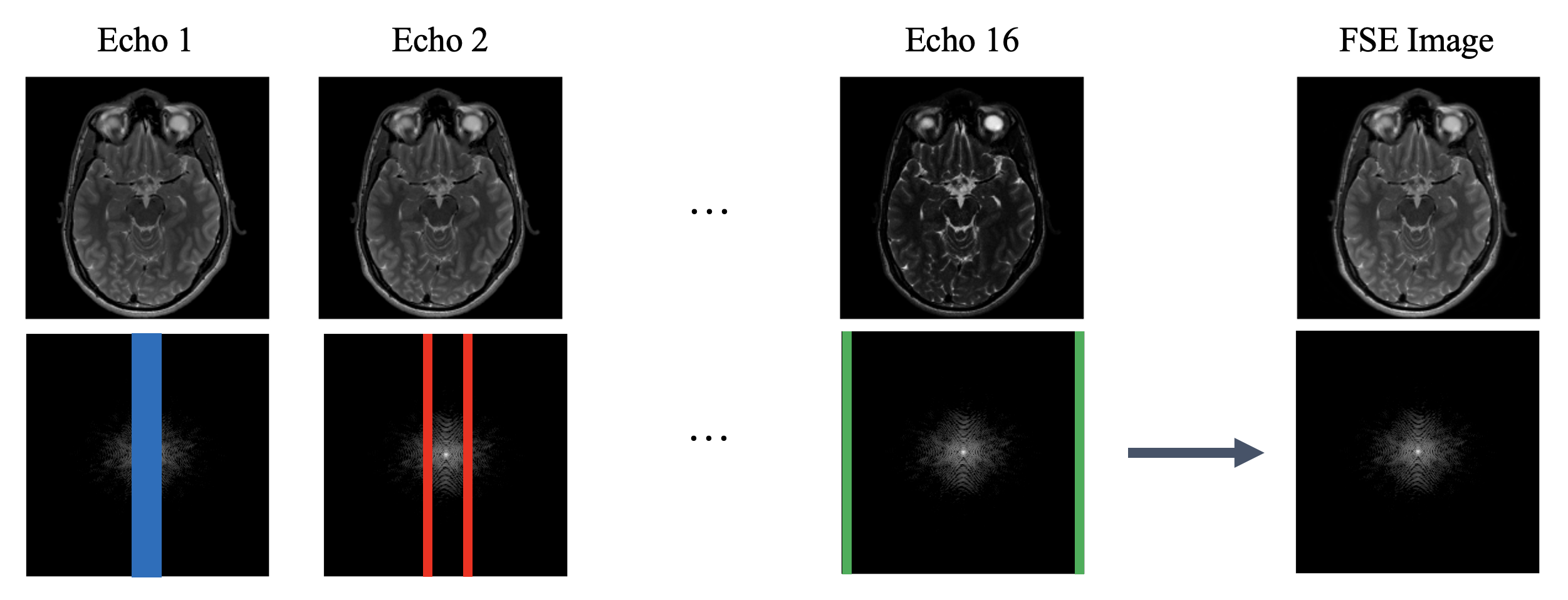}
\caption{Example of how ground truth FSE images were simulated using PD and T2 maps. Top row contains images generated using Eq. \ref{eqn:decay_motion_forward_model} at 16 different echo times with an echo spacing of 12ms. The bottom row shows the respective k-space measurements collected and the sampling pattern used for each echo image to create the FSE image.} \label{fig:FSE_Gen}
\end{figure}

To correctly simulate the FSE sequence acquisition we first estimated T2 and Proton Density (PD) maps from multi-contrast brain MRI scanned at our institution with institutional board approval and informed consent (details about the protocol can be found in the supplementary materials section).
Following the FSE example in \cite{Pulseq}, we simulated echo train lengths (ETL) of 16 and 8, and echo spacing of $12$ms, resulting in $16$ (resp.\ 8) different echo images (Fig. \ref{fig:FSE_Gen}) which depict how the signal changes during an FSE acquisition. To form our ground truth FSE image we used a center-out k-space acquisition ordering along the echo train axis, corresponding to PD contrast (Fig. \ref{fig:FSE_Gen}). The non-overlapping k-space from each echo was combined into one k-space which was inverse Fourier transformed to make the ground truth image. A total of 18 TRs were acquired, corresponding to Nyquist rate sampling.

\subsubsection{FSE Compensated Motion Simulation:}
To demonstrate the importance of accounting for the FSE effects, as proof-of-principle we simulated nine motion events evenly spaced every 2TR seconds. Each rotation angle was randomly chosen from a Normal distribution $\mathcal{N}\left(0, 2\right)$ based on \cite{johnson2019}. We then sampled k-space according to the particular motion state and echo time across the 18 TRs. The IFFT of this k-space resulted in a motion-corrupt image (See Supplementary Material for Simulation Outline). We emphasize that because this motion model used the same phase encode lines for each TR as was used to generate the ground truth image, the simulation correctly takes into account the relationship between the signal decay during the echo trains and the motion state for each TR. 

\subsubsection{FSE Agnostic Motion Simulation}
\begin{figure}
\includegraphics[width=\textwidth]{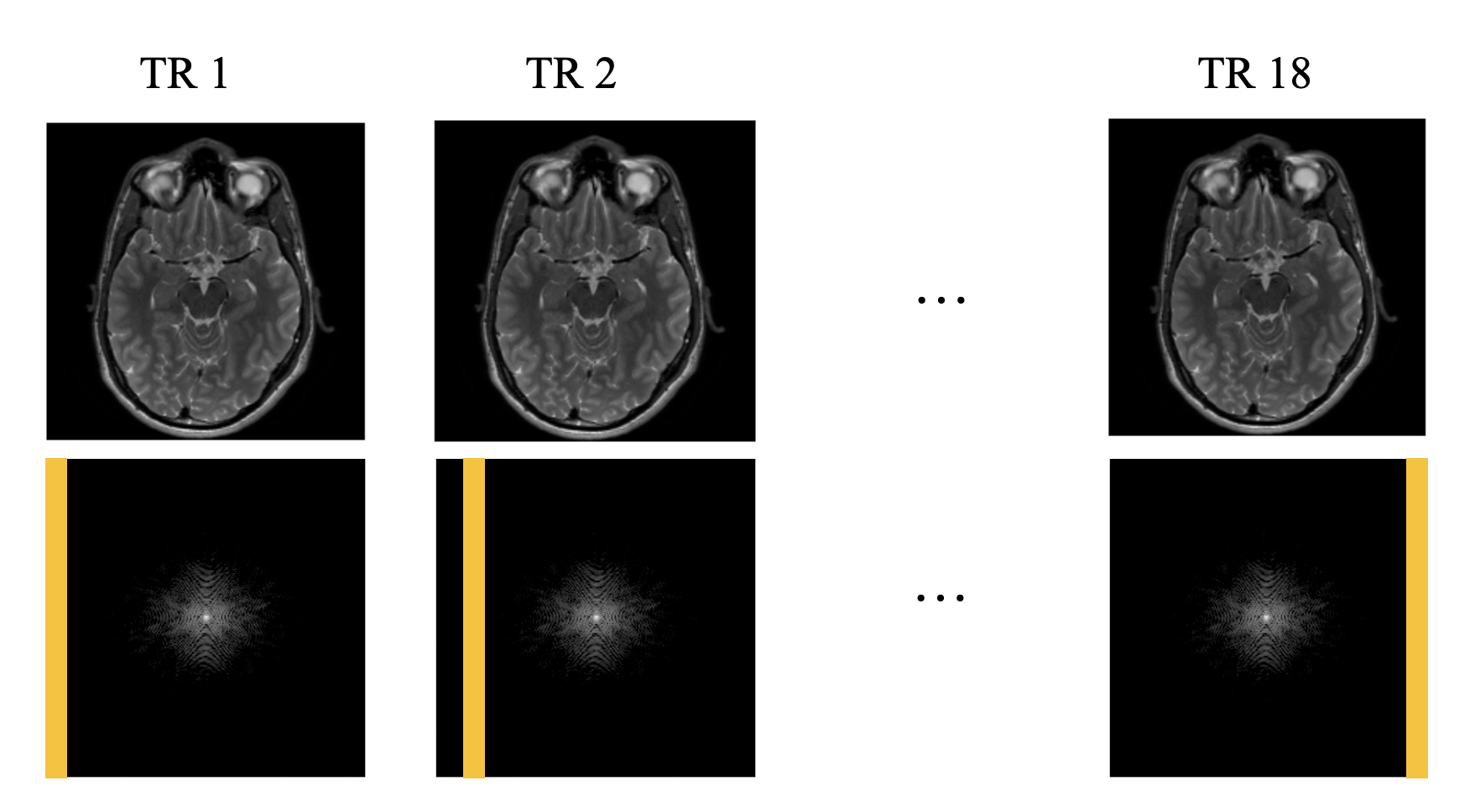}
\caption{Example naive motion-corrupted FSE images. The top row contains images which are rotated version of the GT FSE image. The bottom row depicts the sampling pattern in k-space for each TR. Note that there is no notion of signal decay or echo train ordering.}
\label{fig:FSE_Motion_Gen}
\end{figure}

To generate the FSE agnostic motion-corrupted images we simply took the FSE image (rightmost in Fig. \ref{fig:FSE_Gen}), linearly segmented it into 9 disjoint regions of k-space (corresponding to two TRs each), and applied the same rotations we used previously (shown in Fig. \ref{fig:FSE_Motion_Gen}). Although the angles of rotation remain the same in this method there is no acknowledgment of the temporal relationship found between lines captured in the same echo train for the true FSE acquisition and this is what ultimately causes this method of simulation to diverge from actual FSE scan data.

\subsubsection{Motion Correction Network Training}
To test the effect of the different motion models we trained two conditional GANs: the first model trained with our proposed FSE motion simulation, and the second trained using the FSE agnostic simulation. The input to the conditional GAN was the motion-corrupted image and the output was the motion-free FSE image.
Additional information about the training can be found in the supporting information. Our conditional GAN approach was chosen as it is commonly used as the backbone for deep learning based motion correction methods \cite{johnson2019}. We anticipate that more sophisticated methods would lead to similar overall conclusions.

We implemented image reconstruction using the BART toolbox \cite{bart} and motion correction networks using PyTorch; our code is publicly available\footnote{\url{https://github.com/utcsilab/GAN\_motion\_correction}}. Both models were trained using 819 paired images from the respective motion simulation method. 
We tested both models on data generated by properly simulating an FSE acquisition with motion and tracked both  structural similarity index (SSIM) and normalized root mean square error (NRMSE) values to evaluate image quality and performance.
The results for two different ETL values are summarized in Table \ref{table:error}.

\section{Results and Discussion}
\subsubsection{Results}
\begin{figure}[ht!]
\centering
\includegraphics[width=\textwidth]{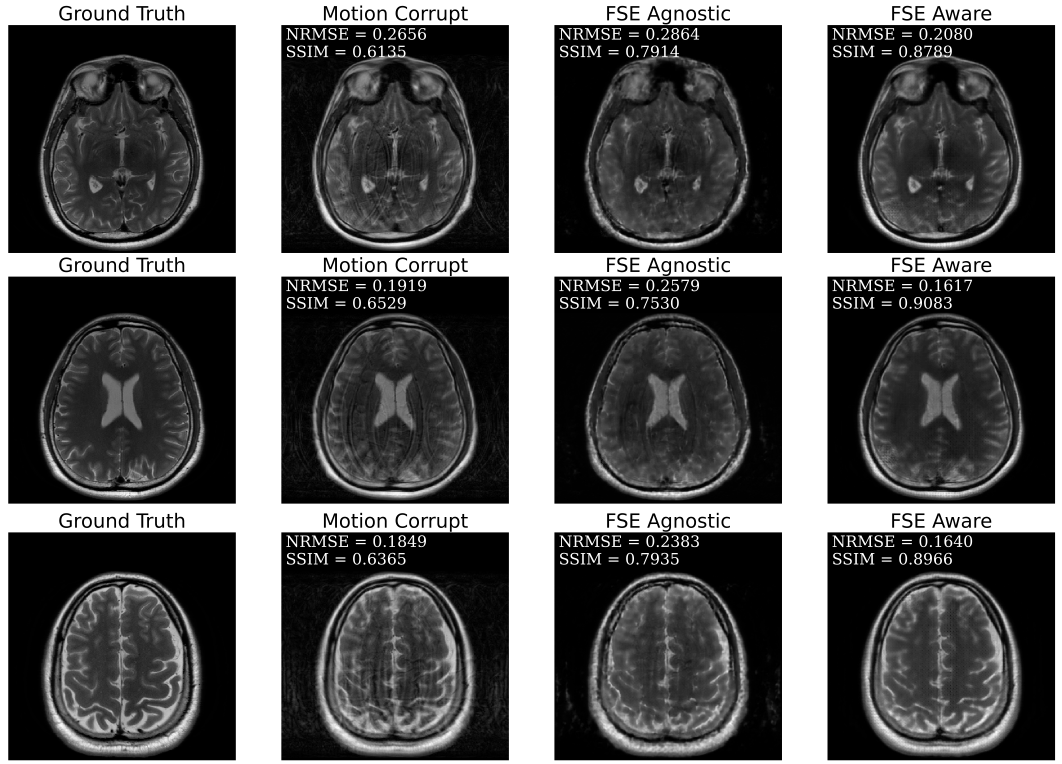}
\caption{Example motion artifact correction for 3 different slices. The first column is the ground truth, second column is for the motion-corrupt images, third column is for motion correction network trained on traditional motion simulation methods and the fourth column shows result for correct physics guided motion simulation methods. } \label{fig:motion_correct_examples}
\end{figure}

Table \ref{table:error} shows the performance of both models when tested on simulated motion-corrupt FSE acquisitions. By training on simulated data which is agnostic to the actual acquisition parameters of an FSE sequence there is a significant drop in quantitative performance when compared to models trained and tested on FSE-aware models for motion-corruption and this was observed for different ETL values.
Fig.~\ref{fig:motion_correct_examples} shows multiple motion correction examples chosen from the test set. All cases show a significant improvement in both quantitative and qualitative metrics when accounting for the FSE acquisition.  The traditional simulation method does not properly account for the correct sampling of k-space, and therefore at test time it is not able to correct the motion artifacts.

\begin{table}[ht]
\centering
\caption{Performance comparison of FSE agnostic vs. FSE aware motion correction on the test set for 2 different ETL values. For the ETL = $16$ case, there are 18 repetition times and time between each echo was $12$ms. Whereas for the ETL = $8$ case, there are $36$ repetition times and time between each echo was $20$ms. For consistency, we simulated $9$ motion states for both the cases. }
\begin{tabular}{|c|c|c||c|c|}
\hline
\textbf{ETL} & 16 & 16 & 8 & 8 \\
\hline
\textbf{Error Metric}                & SSIM                      & NRMSE     & SSIM                      & NRMSE       \\ \hline \hline
\textbf{Input} & 0.554 $\pm 0.071$   & 0.268 $\pm 0.058 $   & 0.659 $\pm 0.053$   & 0.185 $\pm 0.047 $    \\ \hline
\textbf{FSE Agnostic} & 0.730 $\pm 0.060 $  & 0.287 $\pm 0.040 $ & 0.885 $\pm 0.058 $  & 0.120 $\pm 0.025 $ \\ \hline
\textbf{FSE Aware} & \textbf{0.851 $\pm 0.043$}  & \textbf{0.215 $\pm 0.039$} & \textbf{0.939 $\pm 0.022$}  & \textbf{0.119 $\pm 0.016$} \\ \hline
\end{tabular}\label{table:error}
\end{table}

\subsubsection{Discussion}
Many recent works that use deep learning for retrospective motion correction are trained with simulated motion-corruption.  While spin-echo and steady-state sequences are only weakly dependent on the acquisition ordering, FSE acquisitions strongly depend on the data ordering. When this ordering is not correctly accounted for in simulation, there is an expected appreciable drop in performance at test time. This is important to consider as FSE is the backbone of clinical MRI and most publicly available raw MRI datasets are all FSE based. This bias is strongly related to other implicit data crimes arising from the misuse of public data for MRI reconstruction \cite{shimron2021subtle}. Fortunately, this problem is easily mitigated by accounting for the data ordering \cite{MC2-Net}. Another approach to circumventing the issue include unpaired training with experimentally acquired motion-corrupt data \cite{zhu2017unpaired,unpairedMRI}.

Further work is required to investigate the disparity in network performances on prospective motion corrupt data. While our model does not account for through-plane motion in 2D imaging, the model could be extended to 3D FSE acquisitions which have significantly longer ETLs and no through-plane motion.

\section{Conclusion}
When applying data driven methods to retrospective motion correction, simulation parameters are crucial. In particular, when training motion correction networks for FSE sequences the choice of sampling masks across echo trains and repetition times is key to ensuring that the correct artifact distribution is learned by the network to ensure best results at inference on experimentally acquired motion-corrupt data.

\subsubsection{Acknowledgements} 
This work was supported by NIH Grant U24EB029240 and NSF IFML 2019844 Award.

%
%
%

\newpage

\section*{Supplementary Material}

\subsection*{Data Acquisition Protocol}

\subsubsection*{Base Multicontrast Sequence:} The $T_2$ and PD maps used to simulate ground truth images for motion were estimated using a multi delay multi echo (MDME) sequence protocol at our institution with proper IRB approval and informed consent from all volunteers. The protocol consisted of four different inversion times and two different echo times to encode the $T_1$ and $T_2$ information respectively resulting in eight images. For parameter mapping, we simulated a dictionary of signal evolutions matching the sequence parameters at multiple $T_1$ and $T_2$ values and employed dictionary matched filtering by selecting the entry with the highest inner product with the experimentally observed signal. The dictionary was generated by changing $T_1$ values from $100$ to $6000$ms and $T_2$ values from $10$ to $1000$ms with step sizes of $20$ and $2$ms respectively. The MRI scans were conducted on a Siemens 3T Vida scanner (Siemens Healthineers, Erlangen, Germany) and 53 slices were acquired per subject. The $20$ central slices per subject were used for training of the neural network and the remaining slices were discarded. The total duration of the MDME sequence was $8$ minutes with a repetition time (TR) of $7680$ms. The employed echo time values were $27$, $90$ms while the delay times of $7562$, $3504$, $1041$ and $171$ms were used. The acquired frequency matrix was of size $320\times288$ with a field of view of $22.8$cm. The undersampled k-space data was acquired with an acceleration factor of $3.2$ and reconstructed using the BART toolbox \cite{bart}.

\subsection*{Motion Simulation}

Figure 4 shows how the same rigid body motion can induce different motion artifacts based on the acquisition simulation.  Figure 5 outlines the procedure followed to generate FSE aware motion artifacts using PD and $T_2$ maps. 

\begin{figure}[ht!]
\centering
\includegraphics[width=.92\textwidth]{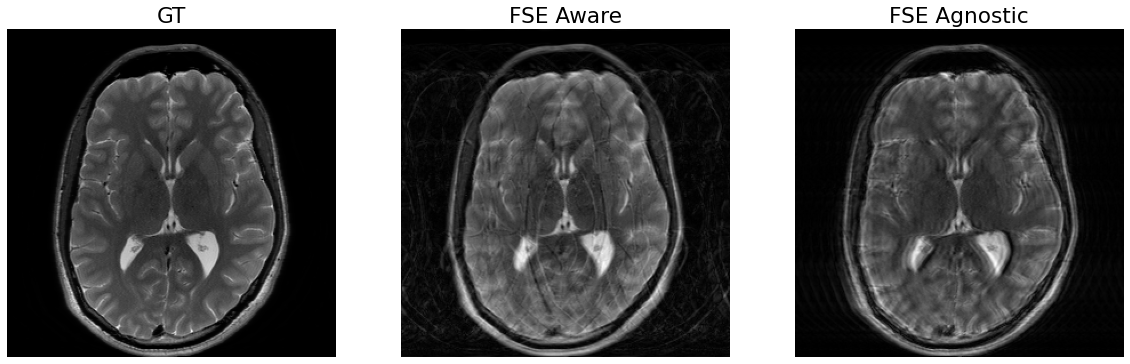}
\caption{Ground truth FSE image (left) with example motion artifacts from FSE Aware simulation (center) and FSE Agnostic simulation (right).} \label{fig:motion_artifact examples}
\end{figure}

\begin{figure}[ht!]
\centering
\includegraphics[width=.58\textwidth]{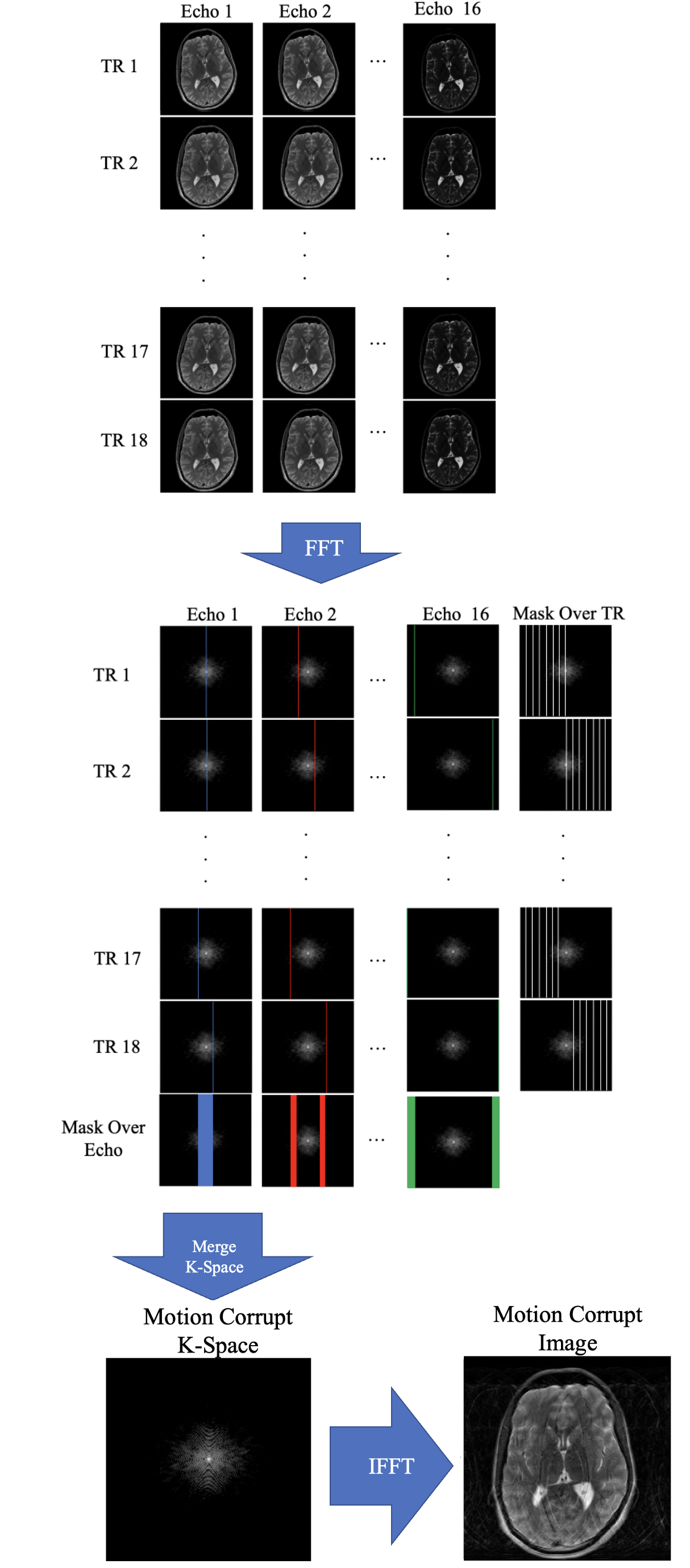}
\caption{Proposed motion simulation for FSE sequences. T2 and PD masks are used to generate 16 images with varying TE. Each echo image is randomly rotated for each TR to simulate inter-TR motion (all echos in the same TR are rotated the same). These images are then transformed into k-space and sampled once (non-overlapping) based on their TR and echo number. All sampled k-space lines are then merged into a single k-space as they would be in a typical FSE acquisition. The IFFT is taken to obtain the motion-corrupt image. } \label{fig:FSE_motion_sim}
\end{figure}

\end{document}